# MicroPython Testbed for Federated Learning Algorithms – The Extended Paper

Miroslav Popovic[a, *], Marko Popovic[b], Ivan Kastelan[a], Miodrag Djukic[a], Ilija Basicevic[a]
[a] University of Novi Sad, Faculty of Technical Sciences, Trg Dositeja Obradovica 6, Novi Sad, Serbia
[b] RT-RK Institute for Computer Based Systems, Narodnog Fronta 23a, Novi Sad, Serbia
[*] Corresponding author: Email address: miroslav.popovic@rt-rk.uns.ac.rs (Miroslav Popovic)

**Abstract**
Recently, Python Testbed for Federated Learning Algorithms emerged as a low code and generative large language models amenable framework for developing decentralized and distributed applications, primarily targeting edge systems, by nonprofessional programmers with the help of emerging artificial intelligence tools. This light framework is written in pure Python to be easy to install and to fit into a small IoT memory. It supports formally verified generic centralized and decentralized federated learning algorithms, as well as the peer-to-peer data exchange used in time division multiplexing communication, and its current main limitation is that all the application instances can run only on a single PC. This paper presents the MicroPyton Testbed for Federated Learning Algorithms, the new framework that overcomes its predecessor's limitation such that individual application instances may run on different network nodes like PCs and IoTs, primarily in edge systems. The new framework carries on the pure Python ideal, is based on asynchronous I/O abstractions, and runs on MicroPython, and therefore is a great match for IoTs and devices in edge systems. The new framework was experimentally validated on a wireless network comprising PCs and Raspberry Pi Pico W boards, by using application examples originally developed for the predecessor framework.

## 1. Introduction

The context of this research is the ongoing EU Horizon 2020 project entitled Trustworthy and Resilient Decentralized Intelligence for edge Systems (TaRDIS) [1]. This project aims to create a toolbox for easy programming of decentralized and distributed applications, primarily in edge systems, such as electrical vehicles charging in a smart grid, privacy-preserving federated learning in smart homes, orbit determination and time synchronization (ODTS) for low Earth orbit (LEO) satellite constellations, and highly resilient factory shop floor digitalization for a factory comprising production lines, a warehouse, and an intralogistics fleet of robots.

Within TaRDIS the Python Testbed for Federated Learning Algorithms (PTB-FLA) [2] is being developed as a framework for federated learning algorithms (FLAs) i.e., as a runtime environment for FLAs under development. PTB-FLA is intentionally written in pure Python to be easy to install (because it has no external dependencies) and to fit to a small IoT memory. It supports formally verified generic centralized and decentralized FLAs [3], as well as the peer-to-peer data exchange used in time division multiplexing (TDM) communication e.g., used in ODTS.

PTB-FLA was designed with a simple API based on a Single Program Multiple Data (SPMD) pattern, to be easily used by ML and AI developers who do not need to be professional programmers, and as such, it is also amenable to generative Large Language Models (LLMs) like ChatGPT. In this respect, recent research advances are twofold. First, the PTB-FLA four-phase development paradigm appeared as a systematic guide for humans to transform a sequential code into a PTB-FLA code [4]. Second, PTB-FLA has been shown to be amenable to AI by successful redevelopment of the three simple examples from [2] using the ad hoc approach, wherein ChatGPT was given the context comprising the input sequential code and was asked to create the target client and server callback functions [5].

Until now, the main PTB-FLA limitation was that all the application instances (i.e., processes) could run only on a single PC (i.e., localhost). This paper presents MicroPyton

Testbed for Federated Learning Algorithms (MPT-FLA), a new FL framework that inherits all the advantages of PTB-FLA while overcoming its main limitations such as the fact that individual application instances may run on different network nodes like PCs and IoTs, primarily in edge systems, but also in the cloud-fog-edge continuum.

Continuing the pure Python ideal, MPT-FLA is based on Python asynchronous I/O (asyncio) abstractions (including asyncio coroutines, streams, and events), and runs on MicroPython (the trimmed-down Python 3 that runs directly on microprocessor/controller hardware and replaces the OS). It is therefore a great match for smart IoTs and devices in edge systems. However, MicroPyton does not support Python multiprocessing abstractions (e.g., process, queue, clients & listeners, etc.) that were the foundation for PTB-FLA, and therefore designing a new MPT-FLA architecture required two evolutionary steps with some tricky design choices, followed by rather involved implementation and validation.

The new framework was experimentally validated on a wireless WiFi network comprising PCs and Raspberry Pi Pico W boards, by using the application examples originally developed for the predecessor framework PTB-FLA adapted here to the new MPT-FLA application programming interface (API). This experimental validation was successful, as the MPT-FLA adapted application examples produced the same numerical results as the originals running on PTB-FLA.

The main original paper contributions are: (1) a novel FL framework dubbed MPT-FLA which is based on MicroPython and Python asyncio, (2) a new set of application examples that are adapted to the new MPT-FLA API, and (3) the experimental validation approach, results, and discussion, which may be useful to other researchers.

The rest of the paper is organized as follows. Section 1.1 presents closely related work, Section 2 presents the necessary materials, Section 3 presents the MPT-FLA design, Section 4 presents the MPT-FLA experimental validation with results and discussion, and Section 5 concludes the paper.

*1.1. Related work*

At present, the most prominent federated learning (FL) frameworks, namely TensorFlow Federated (TFF) [6], [7] and BlueFog [8] work well in cloud-edge continuum. However, they are not deployable to edge only, they are not supported on OS Windows, and they have numerous dependencies that make their installation complex and time-consuming. A comprehensive comparative review and analysis of open-source FL frameworks for IoT, made recently in 2021 by Kholod et al. [9], indicates that developing an FL framework targeting edge systems is still an open challenge.

MPT-FLA was already introduced in Section 1, and at this point, it seems appropriate to clarify what it is and what it is not. MPT-FLA is primarily an FL framework, which is seen by ML and AI developers in the TaRDIS project as an "algorithmic" testbed where they can plug in and test their FLAs. MPT-FLA is neither a complete system such as CoLearn [10] and FedIoT [11] nor a system testbed such as the one that was used for testing the system based on PySyft in [12].

The MPT-FLA programming model is based on the SPMD pattern [13] like other well-known programming models: MapReduce, MPI, OpenMP, and OpenCL. Interestingly, MPT-FLA and MapReduce are rather similar – the client callback function in MPT-FLA is like the map callback function in MapReduce, whereas the server callback function in MPT-FLA is like the reduce callback function in MapReduce.

A possible topic of our future work is to try to adapt MPT-FLA for Codon [14], a compiler for high-performance Pythonic applications and Domain Specific Languages (DSLs). In a broader sense, Codon is a full language (and compiler) that borrows Python 3's syntax and semantics and compiles to native machine code with low runtime overhead, allowing it to rival C/C++ in performance. Codon uses a novel bidirectional Intermediate Representation (IR) and compiles it to LLVM IR that is in Static Single-Assignment (SSA) form (SSA-form), which enables code analysis including code formal verification like in [15].

## 2. Materials and methods

The experimental WiFi network that was used for the MPT-FLA experimental evaluation consists of one WiFi router Belkin F5D7234-4, two Raspberry Pi Pico W boards, and one PC. The next subsections specify each component type, respectively.

*2.1. WiFi router Belkin F5D7234-4*

**Hardware**. The "Wireless G Router" from Belkin is an Internet router with a standard 802.11g wireless interface. The router allows Internet connection sharing among computers connected to a local area network (LAN). Compatible with the 802.11g and 802.11b protocols, the router is designed with maximum performance and compatibility in mind [19]. When used with other wireless devices, the router supports transfer rates of up to 54Mbps.

*2.2. Raspberry Pi Pico W boards*

**Hardware**. Raspberry Pi Pico W is a low-cost microcontroller board based on the RP2040 chip enabling IoT-embedded electronics projects. The onboard 2.4GHz wireless interface (802.11n) is a single-band (2.4 GHz) WiFi with WPA3 and Soft Access Point (SAP) supporting up to 4 clients. For best wireless performance, the antenna should be in free space (adding grounded metal to the sides of the antenna can improve the antenna's bandwidth). Inside the RP2040 is a "permanent ROM" USB UF2 bootloader, which is used to program the new firmware.

The RP2040 has 264KB of onboard RAM, and instead of a built-in flash memory, it uses the external QSPI flash chip. The 2MB on this board is shared between the program running at the time and any file storage used by MicroPython. The main RP2040 chip features include (1) dual ARM Cortex-M0+ @ 133MHz, (2) 264kB on-chip SRAM, (3) support for up to 16MB of off-chip flash memory, (4) 30 GPIO pins, and (5) peripherals including 2 UARTs, 2 SPIs, 2 I2Cs, 16 PWM channels, USB 1.1 controller and PHY, etc.

**Firmware**. We used the firmware file: "RPI_PICO_W-20231005-v1.21.0.uf2", which essentially contains the official MicroPython port for Raspberry Pi.

*2.3. PC*

**Hardware**. Dell Latitude 3420 laptop. Processor: 11th Generation Intel® Core™ i7-1165G7 (12 MB Cache, 4 Core, 8 Threads, 2.80 GHz to 4.70 GHz, 15 W). Installed RAM: 16.0 GB (15.7 GB usable). System type: 64-bit operating system, x64-based processor.

**Software**. OS Windows specifications: Edition Windows 10 Pro. Python 3.12.0 on win32. Notepad++ version 8.6. The Integrated Development Environment (IDE) VS Code Version: 1.85.2.

**MPT-FLA**. By the end of Q2 2024, MPT-FLA should become publicly available in the GitHub repository: https://github.com/miroslav-popovic/ptbfla.

## 3. MPT-FLA design

Here we present a short history of MPT-FLA evolution. The predecessor PTB-FLA framework [2] was based on the following Python multiprocessing abstractions (i.e., classes): the process, the queue, and the clients & listeners. However, MicroPython does not support these abstractions (the Python multiprocessing package port is not provided), and consequently, PTB-FLA cannot run on MicroPython.

Briefly summarizing our development effort, we may say that the MPT-FLA framework evolved from its predecessor PTB-FLA framework in two evolutionary steps. In the first step, which was not completely successful, and therefore should be considered as an intermediate step, we replaced the Python multiprocessing abstractions with somewhat similar abstractions available in MicroPython in the following way: (1) the local server process was replaced with the local server thread, (2) the queue was replaced with the list (because the queue for threads is also not supported in MicroPython) and the spin with the sleep when receiving messages from the list, and (3) the clients & listeners were replaced with sockets. The resulting architecture, with sleeps in order of tens of ms, was functional and acceptable for PCs but not for Raspberry Pi Pico W boards where such short sleeps are problematic. On the other hand, increasing sleeps to the order of seconds yields poor performance.

In the second step, we introduced the appropriate Python asyncio abstractions as follows. The sockets were replaced with the asyncio streams, the threads were replaced with the asyncio coroutines, and the spin with the sleep when receiving messages was replaced with the blocking wait on the asyncio event (the asyncio event was the best possible replacement because MicroPython does not support asyncio queue). The resulting architecture, dubbed MPT-FLA architecture, can directly run both on regular Python on PCs and on MicroPython on Raspberry Pi Pico W boards and has acceptable performance on both PCs and Raspberry Pi Pico W boards. The following subsections present the MPT-FLA system architecture, the MPT-FLA API, and the MPT-FLA operation, respectively.

*3.1. MPT-FLA system architecture*

The MPT-FLA-based system, or more briefly the MPT-FLA system, is an edge system that may be represented as a graph with $n$ nodes (with the node identifications from 1 to $n$, respectively; note: in Python encoded from 0 to $n - 1$), which are interconnected with edges, where the nodes are processes hosted by smaller subsystems such as IoTs, smart devices, computers, etc., and the edges are communication links (TCP connections) over the Internet edge network, see Fig. 1. The edges are shown as dashed lines because the corresponding communication links are dynamically created on demand to allow message exchange between the peers and are destroyed afterward.

At the system startup, and during most of the time, no links exit, and during system evolution (i.e., execution) only the required links will be dynamically created (and destroyed). Moreover, the links that are never used will never be created e.g., if $n_i$ and $n_j$ never communicate the link between them will never be created. The set of links that will actually be created (and used, and destroyed), depends on the system operation mode i.e., the generic distributed and decentralized algorithm executed by MPT-FLA. When visualizing only the links that are actually created, the general graph shown in Fig. 1 takes various forms: in the case of the generic centralized FLA it takes the form of a star (with a star being the FLA server), in the case of the generic decentralized FLA it takes the form of a clique (fully connected graph), and in the case of the TDM peer data exchange (also called the TDM communication round), it takes the form of an arbitrary graph (wherein only the pairs of nodes that communicate are connected).

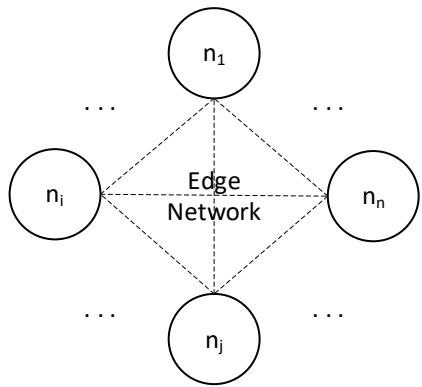

Fig. 1. The MPT-FLA system architecture.

Generally, a node $n_i$ is a process executing an application instances $a_i$. Each $a_i$ initially creates its testbed instance $t_i$. A set of all the application instances constitutes a distributed application $A = \{a_1, \ldots, a_n\}$, whereas a set of all the testbed instances constitutes a distributed testbed $T = \{t_1, \ldots, t_n\}$. During normal system operation, the distributed application $A$ uses the distributed testbed $T$ to execute the desired distributed algorithm by using the appropriate generic distributed algorithm offered by MPT-FLA. Currently, MPT-FLA supports three generic distributed algorithms: (1) the generic centralized FLA, (2) the generic decentralized FLA, and (3) the TDM peer data exchange algorithm.

In this paper, nodes are processes hosted by either Raspberry Pi Pico W boards (briefly Pico boards) with MicroPython or by PCs with OS Windows (or Ubuntu) and regular Python. Normally, one Pico board hosts a single process (node), whereas one PC may host one or more processes (nodes). A process at a Pico board is automatically started by MicroPython when the Pico board is powered on (MicroPython starts executing the module main.py at the root of the file system), whereas a process at a PC is started by a special launcher process $s$, the latter is typically manually started from the command line interface.

The MPT-FLA system architecture is hierarchically organized in the following three layers (see Fig. 2): the Application layer (including the script launch on PCs or the module main on Pico boards and the application modules on both), the MPT-FLA layer (including the module launcher on PCs or config on Pico boards and the modules mp_async_ptbfla and mp_async_mpapi on both), and the Python layer (including the packages asyncio and subprocess). The tokens "mp" and "async" in the prefix "mp_async" stand for MicroPython and asyncio, respectively. Besides the horizontal layers, Fig. 2 is organized in three vertical lanes showing modules that are used for startup on PCs (the left lane), for startup on Pico boards (the middle lane), and for normal operation on both (the right lane), respectively.

Next, we proceed by the lanes, from the left to the right, to explain the startup on PCs, the startup on Pico boards, and the normal operation on both.

**Startup on PCs**. The script launch relies on the following conventions. The first two command line arguments of the application instance $a_i$ are: the number of nodes $n$ and the node identification (ID) $i$ (the rest of the arguments are application specific). The first three command line arguments of the script launch are: the name of the application's main Python module (e.g. example1.py), the number of nodes $n$, and the node ID specifier (the rest of the arguments are just passed to the individual application instances). The node ID specifier is either "id" (meaning all the IDs) or "i-j" (meaning the IDs in the interval $[i, j]$, where $i$ and $j$ are in $[0, n–1]$ and $j >= i$). Based on these conventions, the script launch creates command line arguments of individual application instances by mapping the node ID specifier into the corresponding IDs.

Depending on the node ID specifier, the script launch launches: (i) a single node $n_i$ (if the node specifier is "i-i"), (ii) a group of nodes from $n_i$ to $n_j$ with IDs in the interval $[i, j]$ (if the node ID specifier is "i-j" and $j > i$), or (iii) all the nodes from $n_0$ to $n_{n-1}$ (if the node ID specifier is "id"). The script launch uses the module launcher, which in turn uses the class Popen from the package subprocess. The launched nodes (processes) execute in separate terminals (i.e., windows). Note that the script launch may be used on more PCs to start individual or groups of nodes on them.

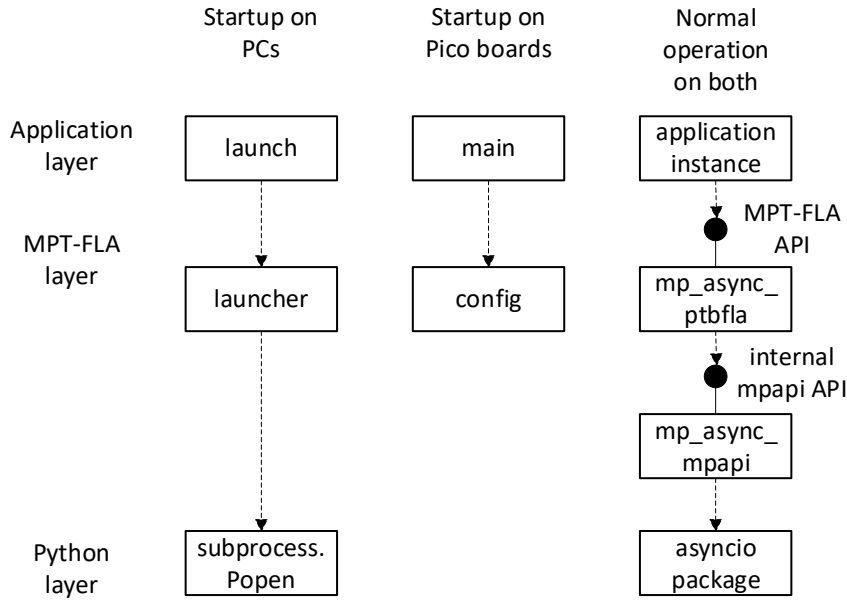

Fig. 2. The UML class diagram of the MPT-FLA system architecture.

**Startup on Pico Boards**. The modules main and config are used for startup on Pico boards – the former plays the role of a launcher, whereas the latter contains the node's configuration that is stored in the dictionary *cfg* (including the WiFi SSID, WiFi password, the command line arguments vector *argv* for this application instance, and the IP address for this node – all these parameters must be set beforehand except the IP address which is set at the node startup).

On power on, MicroPython starts the main thread, which executes the module main, which in turn connects to the specified WiFi network, copies the *argv* from the module config to the *argv* of the main thread (i.e., to the sys.argv), and calls the function run_main in the main application module. (On both Pico boards and PCs, the function run_main creates the async event loop and runs the main async coroutine.)

**Normal operation on both PCs and Pico Boards**. The application instance uses the MPT-FLA API provided by the class PtbFla, within the module mp_async_ptbfla, to create and destroy a testbed instance (by calling the constructor and distractor, respectively), to start up this node (by awaiting the coroutine start), and to execute one of the three generic distributed algorithms (by awaiting the coroutines fl_centralized, fl_decentralized, or get1Meas, respectively) – more details on this API are provided in Section 3.2.

The class PtbFla uses the internal mpapi API (mpapi is the abbreviation of the term "message passing API"). The mpapi API provides the following five asyncio coroutines: (1) server_fun is executed by the asyncuo task, which is created by the PtbFla coroutine start, (2) sendMsg sends a single message, (3) rcvMsg receives a message, (4) broadcastMsg broadcasts a message, and (5) rcvMsgs receives a given number (typically $n–1$) of messages.

Note: The mpapi API is strictly an internal API providing services to the class PtbFla only, and it should never be used by the distributed algorithms developers, instead they should only use the MPT-FLA API in their application program modules.

*3.2. MPT-FLA API*

The MPT-FLA API comprises the following six members (the first is a function, whereas the next four are coroutines declared as such by the keyword async):
1. None PtbFla(*noNodes*, *nodeId*, *flSrvId*=0, *mrIpAdr*='localhost')
2. None async start()
3. *ret* async fl_centralized(*sfun*, *cfun*, *ldata*, *pdata*, *noIters*=1)

4. *ret* async fl_decentralized(*sfun*, *cfun*, *ldata*, *pdata*, *noIters*=1)
5. *obs* async get1Meas(*peerId*, *odata*)

The first member is the constructor with the following arguments: *noNodes* is the number of nodes (or processes), *nodeId* is the node identification, *flSrvId* is the server id (default is 0; this argument is used by the coroutine fl_centralized), and *mrIpAdr* is the master node IP address that must be known by all the nodes. (Note that the destructor is not explicitly enlisted above because it is implicitly called by the garbage collector.)

The second member is the coroutine start, which starts up this node by conducting its role in the complete system startup, see Section 3.3.1.

The third and the fourth members are the generic centralized FLA and the generic decentralized FLA, respectively. By awaiting this coroutine, this node takes its part in the whole distributed algorithm, see Sections 3.3.2 and 3.3.3. The arguments of these two coroutines are: *sfun* is the server callback function, *cfun* is the client callback function, *ldata* is the initial local data (typically a machine learning model), *pdata* is the private data (typically a training data that is used to train the model), and *noIters* is the number of iterations that is by default equal to 1 (for the so called one-shot algorithms), i.e., if the calling function does not specify it, it will be internally set to 1. The return value *ret* is the node final local data. Generally, all data (*ldata*, *pdata*, and *ret*) is application specific.

The fifth member is the TDM peer data exchange algorithm. By awaiting this coroutine, this node exchanges its data with its peer in the current TDM time slot, see Section 3.3.4. The arguments of this coroutine are: *peerId* is the peer node ID, and *odata* is the data of this node to be sent to the peer. The return value *obs* is the data received from the peer.

*3.3. MPT-FLA operation*

This section provides an overview of the MPT-FLA operation by presenting the following four most important scenarios, which correspond to the MPT-FLA main operating modes: (i) the system startup, (ii) the generic centralized FLA execution, (iii) the generic decentralized FLA execution, and (iv) the TDM peer data exchange.

*3.3.1. System startup*

MPT-FLA relies on the following conventions. The node $n_1$ is the master node and its IP address is known to all the nodes (e.g., in case of Pico boards it may be stored as an element of the variable argv in the module config). The TCP port number for the node $n_i$, $pt_i$, is defined as $pt_i = 6000 + i$. Note: conceptually, the index $i$ takes values from 1 to $n$, whereas in MPT-FLA Python implementation it goes from 0 to $(n-1)$.

The system startup handshake has two phases, see Fig. 3. In the first phase, the instance $a_1$ (at the master node) is receiving $(n-1)$ initial messages from all other instances, where the initial message from the instance $a_i$, $i = 2, …, n$, is the list [$id_i$, $ip_i$, $pt_i$] and $id_i$ is the node $n_i$ ID, $ip_i$ is the node $n_i$ IP address, and $pt_i$ is the node $n_i$ port number.

In the second phase, instance $a_1$ broadcasts the final message to all other instances. The final message $f$ is the list of all the lists received by $a_1$ extended with the list for $a_1$ and sorted by $id_i$ i.e., $f = [[id_1, ip_1, pt_1], … [id_i, ip_i, pt_i], … [id_n, ip_n, pt_n]]$ (note: to keep Fig. 3 more readable, $f$ is shown just with the middle list).

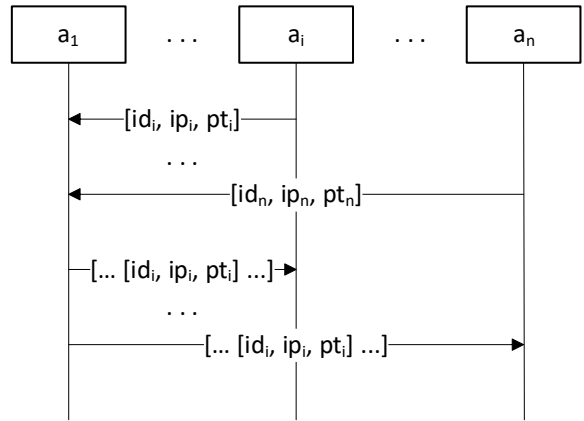

Fig. 3. The system startup.

*3.3.2. Generic centralized FLA execution*

Assume that the instance $a_1$ is the FL server and the other instances $a_i$, $i = 2, \ldots, n$, are the FL clients. Both the server and the clients are taking their part in the collective execution of the generic centralized FLA by awaiting the coroutine fl_centralized on their testbed instances.

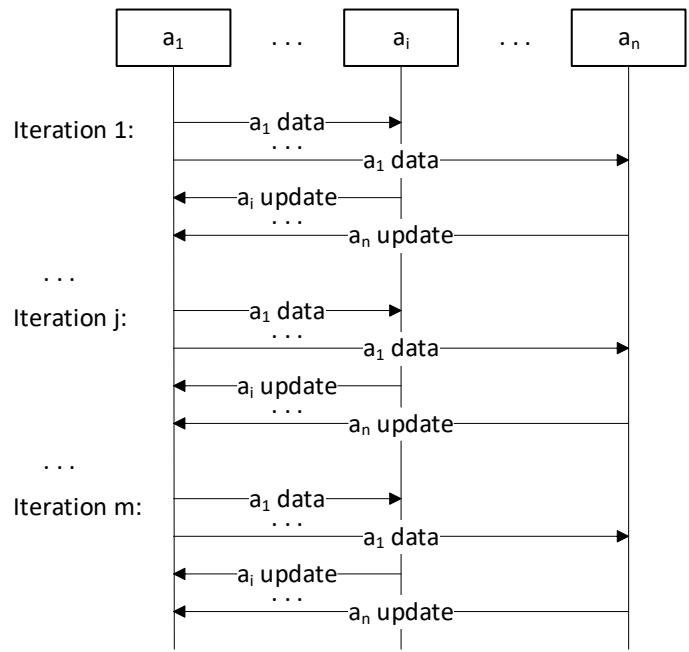

Fig. 4. The generic centralized FLA execution.

The generic centralized FLA has *m* iterations, see Fig. 4, and each iteration has the following three phases.

**Phase 1**. The server broadcasts its local data ("data" in Fig. 4) to the clients, which in their turn call their callback function to get the update data and store the update data locally.

**Phase 2**. The server receives the update data ("update" in Fig. 4) from all the clients.

**Phase 3**. The server calls its callback function to get its update data (e.g., aggregated data), and stores it locally.

After the last iteration, each instance gets its final local data as the return value from the coroutine fl_centralized.

*3.3.3. Generic decentralized FLA execution*

All the instances are taking their part in the collective execution of the generic decentralized FLA by awaiting the coroutine fl_decentralized on their testbed instances. The generic decentralized FLA has *m* iterations, and each iteration has three phases, see Fig. 5, which for its clarity shows just the iteration *j*. Here all the instances are equal peers, and in each iteration, they all periodically switch roles from the server in Phase 1, to the client in Phase 2, and back to the server in Phase 3.

These three phases have the same basic meaning (semantics) as in the generic centralized FLA: in Phase 1, instance broadcasts its local data, in Phase 2, it replies to each Phase 1 message it received by sending its local data update, and in Phase 3, it aggregates all Phase 2 messages it received. For easier comprehension, Fig. 5 shows the simplest scenario where all the messages are sent instantaneously (with zero delay) and therefore are not interleaved.

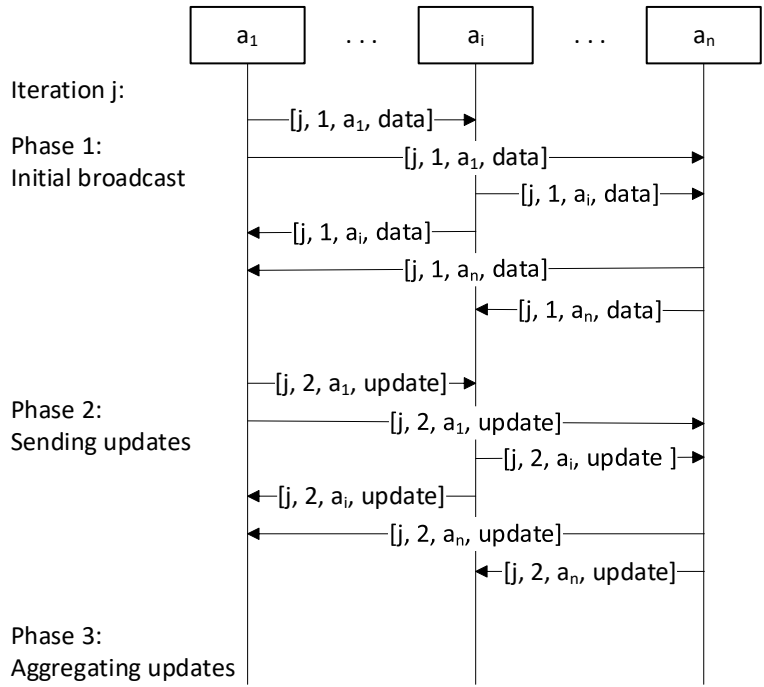

Fig. 5. The generic decentralized FLA execution.

However, unlike the generic centralized FLA that uses simple messages comprising only data (local or update), the generic decentralized FLA, being more complicated, uses messages that are lists which include the following elements: the iteration number, the phase number, the message source address (i.e., the source network address, which is the tuple made of the IP address and the TCP port number, which is denoted as $a_i$ in Fig. 5), as well as the data (local or update).

The second major difference between the generic centralized and decentralized FLAs is related to the message order on the receiving side. Since messages are sent asynchronously, they may be interleaved in both algorithms, but in the case of the generic centralized FLA, they will always be received in both the iteration order and the phase order, whereas in the case of the generic decentralized FLA, they may be received both out of the iteration order and the phase order.

More precisely, on the slower instances, during the whole iteration *j*, some messages may be received from Phase 1 of the next iteration *j* + 1, and during Phase 1 (of iteration *j*), some messages may be received from Phase 2 (of iteration *j*). To postpone their processing to the right iteration and the right phase, the algorithm (i.e., the instance executing it) stores the former messages into the FIFO buffer $buf_1$, and the data from the latter messages into the FIFO buffer $buf_2$. (Note: $buf_1$ stores complete messages whereas $buf_2$ stores data from the messages.)

After these introductory notes, the three phases (in iteration $j$) are summarized in more detail below.

**Phase 1**. An instance acts as a server, and it sends its local data to all its neighbours. These messages have the phase number 1, and each instance sends $(n - 1)$ such messages. Note that each instance is also the destination for $(n - 1)$ such messages.

**Phase 2**. This phase has two subphases, and an instance acts as a client in both. In Subphase 2.1, an instance drains the buffer $buf_1$ i.e., it gets messages from the top of $buf_1$, one by one, and for each one of them, it calls the client callback function to get the update data, and then sends the reply to the message source.

In Subphase 2.2, an instance may receive either a message from the next iteration $(j + 1)$ or from this iteration $j$.

If the message is from the iteration $(j + 1)$, an instance just stores it in the buffer $buf_1$ for later processing in the iteration $(j + 1)$.

If the message is from this iteration $j$, it may be either from Phase 1 or Phase 2. If the message is from Phase 2, an instance just stores it in the buffer $buf_2$ for later processing in Phase 3, whereas if the received message is from Phase 1, an instance calls the client callback function to get the update data, and then sends the reply to the message source.

Note that during the whole Phase 2, an instance does not update its local data, it just passes the update data it got from the client callback function to the message source.

Phase 2 is completed after an instance has received and processed all $2(n - 1)$ messages (from both Phase 1 and Phase 2 of iteration $j$).

**Phase 3**. An instance again acts as a server, it calls the server callback function (and passes it the buffer $buf_2$) to get its update data (e.g., aggregated data), and stores it locally.

Finally, after the last iteration, each instance gets its final local data as the return value from the coroutine fl_decentralized.

*3.3.4. TDM peer data exchange*

TDM communication is organized in a periodic series of rounds, also called a block (or frame) of time slots. Depending on the space environment and communication infrastructure, a single or multiple communications may take place within a time slot. TDM communication is used by various algorithms, and one prominent example is the distributed Orbit Determination and Time Synchronization (ODTS) algorithm used by a constellation of satellites, also known as Space Vehicles (SVs), together with fixed Ground Satiations (GSs). During an ODTS time slot, pairs of peer SVs that have a clear line of sight (LoS) exchange their data i.e., their orbital data (abbreviated as odata).

To support the TDM communication, the MPT-FLA provides the TDM peer data exchange algorithm that is implemented as the coroutine get1Meas, which basically exchanges the caller data with its peer data in the current time slot. The messages that are used for this exchange are the lists with the following elements: the time slot number, the node ID, and the data.

All the instances that take part in the collective TDM data exchange in the current time slot, do this by awaiting the coroutine get1Meas on their testbed instances. Fig. 6 shows an example where the pairs $(a_1, a_j)$ and $(a_i, a_n)$ exchanged their *odata* during the time slot $k$.

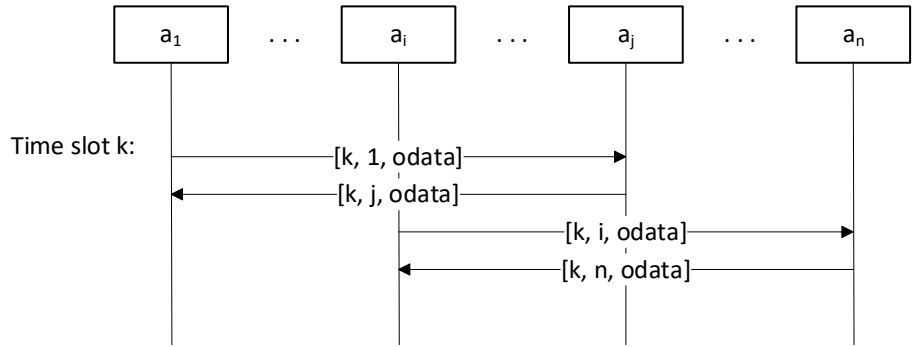

Fig. 6. The TDM peer data exchange.

In the TDM communication of real SVs, the SVs first adjust their antennas to make the communication possible, and then use half-duplex communication to exchange data. Therefore, the messages received from other SVs are not interleaved. However, since instances are executed as processes on PCs (or Pico boards), these messages are delayed and interleaved, and for that reason get1Meas needs to reorder messages to simulate the reality.

For reordering messages, get1Meas maintains the current time slot (initially set to 0) and the buffer of messages sent to this instance, which is implemented as the dictionary of key-value pairs, where the key is the time slot number, and the value is the message. These are testbed internal data.

The two important assumptions that clearly define the contract between get1Meas and its user, are the following: (a) if an instance takes part in communication in the given time slot, it can talk just to one other node in that time slot, (b) if an instance does not take part in communication in a time slot, then it should skip that time slot by awaiting get1Meas and setting *odata* to None.

The TDM peer data exchange algorithm comprises the following five steps.

**Step 1**. If *odata* is None, increment the current time slot and return None (this step supports skipping the current time slot).

**Step 2**. Send the message (carrying this instance data) to the peer.

**Step 3**. If the message for the current time slot is in the buffer of received messages, extract it (get from and delete in) from the buffer, and skip Step 4.

**Step 4**. In the infinite loop, receive the next message sent to this instance. If the next message is not for the current time slot, store it in the buffer, and continue the loop, otherwise break the loop, and proceed to Step 4.

**Step 5**. Increment the current time slot and return the peer data.

**4. MPT-FLA experimental validation – results and discussion**

The MPT-FLA framework was experimentally validated on the experimental WiFi network, consisting of one WiFi router Belkin F5D7234-4, two Raspberry Pi Pico W boards, and one PC (as specified in Section 2), by using adapted algorithm examples that were originally developed for the PTB-FLA framework. The MPT-FLA successfully completed this experimental validation because, as expected, the adapted algorithms produced the same numerical results as the originals, and this was the sole goal of this experiment validation. Since the MPT-FLA is still under development it was too early to evaluate its performance by measuring the metrics such as execution time, communication latency, energy consumption, etc. and this is an important item for our future work.

The next subsections present the adapted algorithms with their specific configurations and operational instructions. The last subsection, Subsection 4.5, provides the overall discussion of the experimental evaluation.

*4.1. Example 1: Federated map*

This example is in fact the adapted Example 1 from [2]. Its purpose is averaging the number of sensor readings above the given threshold, and its pseudocode is given in Table 1 (when compared with [2] the new or modified lines are the lines 1, 3, 4, 7, 9, 11, 13, and 16-18). For the sake of completeness of this paper, we briefly explain all the new pseudocodes.

Table 1
Example 1 pseudocode.

```
01: example1(noNodes, nodeId, flSrvId, mrIpAdr)
02:   // Create PtbFla object and await the system startup
03:   ptb = PtbFla(noNodes, nodeId, flSrvId, mrIpAdr)
04:   await ptb.start()
05:   // Set localData for FL server/clients as follows
06:   if nodeId == flSrvId then
07:      localData = 20.83 // Set the threshold
08:   else   // Set the client readings (for nodes on PCs)
09:      localData = 20.0
10:      if nodeId == noNodes – 1 then
11:         localData = 21.39
12:   // Await fl_centralized with noIterations = 1 (default)
13:   ret = await ptb.fl_centralized(servercb, clientcb, localData, None)
14: clientcb(localData, privateData, msg)
15:   clientReading = localData
16:   if sys.platform == 'rp2' then
17:      reading = sensor_temp.read_u16() * (3.3 / 65535)
18:      clientReading = 27 - (reading - 0.706) / 0.001721
19:   threshold = msg
20:   tmp = 0.0
21:   if clientReading > threshold then
22:      tmp = 1.0
23:   return tmp
24: servercb(privateData, msgs)
25:   listOfIsOverAsFloat = msgs
26:   return sum(listOfIsOverAsFloat) / len(listOfIsOverAsFloat)
```

The main functions of Example 1: (i) creates the testbed object *ptb* and awaits the system startup, (ii) sets the instance initial local data depending on its *nodeId*, and (iii) awaits the API coroutine fl_centralized on *ptb*. The initial local data for the server is the temperature threshold 20.83 °C i.e., 69.5 °F, whereas the initial local data for the clients are their simulated (i.e., default) sensor readings, which have the value 20.0 °C i.e., 68.0 °F (below the threshold) for all the clients except the last one whose reading is 21.39 °C i.e., 70.5 °F (above the threshold). Note that for the instances running on PCs the simulated sensor readings will remain unchanged, whereas for the instances running on Pico boards the sensor readings will be overwritten by the readings from real temperature sensors.

The client callback function clientcb: (i) receives the client local data (its default sensor reading) through the argument *localData* and the server local data (the threshold) through the argument *msg* (the message from the server), (ii) if the function runs on the Pico board (i.e., sys.platform is set to 'rp2') then it reads the real sensor, converts the value from °F to °C and overwrites the default reading (see lines 16-18), (iii) sets the variable *tmp* to 0.0 if the reading is below the threshold or to 1.0 otherwise, and (iv) returns *tmp* to the generic coroutine fl_centralized, which in turn forwards the *tmp* to the server. The server in turn collects all the client replies into a list and passes this list to the server callback function.

The server callback function servercb receives this list through the argument *msgs*, and in turn returns the fraction of sensor readings that are above the threshold.

The system configuration (comprising the configuration parameters of the individual instances) for the experimental validation is given in Table 2. The node index for the node $n_i$ has the value $(i – 1)$. The node indexes are the internal encodings of node IDs used in Python (program code and arguments). Note that the last argument '192.168.2.4' was the specific IP address of the master node (when repeating the experimental validation this might be different and then should be modified accordingly).

Table 2
Example 1 system configuration.

| Node index | Hosted by | Argument vector (argv) for the node |
|---|---|---|
| 0 | PC | ['mp_async_example1_fedd_mean.py', '3', '0', '0', '192.168.2.4'] |
| 1 | Pico board | ['mp_async_example1_fedd_mean.py', '3', '1', '0', '192.168.2.4'] |
| 2 | Pico board | ['mp_async_example1_fedd_mean.py', '3', '2', '0', '192.168.2.4'] |

The operational instructions for the experimental validation are given in Table 3. The abbreviation rp2 stands for the Raspberry Pi Pico W board. The module main should import the python module "mp_async_example1_fedd_mean.py".

In case when the experiment does not depend on the environment, the expected results may be provided beforehand by running all the instances on the PC using the script launch (the results for individual nodes are then displayed in separate terminals, respectively). However, Example 1 depends on the real temperatures at places where the Pico boards are situated. Under the assumption that the equipment used in the experiment is in good order, there are three possible cases for the results at the server. If both temperatures are below (above) the threshold, both client instances should return 0.0 (1.0) and the result evaluated by the server instance should be 0.0 (1.0). The third case is when one of the temperatures is below and the other is above the threshold, then the corresponding client instances should return 0.0 and 1.0, respectively, and the result evaluated by the server instance should be 0.5.

Table 3
Example 1 operational instructions.

| Step no. | Step | Description (comment) |
|---|---|---|
| 1 | Prepare rp2 no. 1 | Connect rp2 to PC USB. Start VSCode. Edit config.py & main.py. Upload project to rp2. Disconnect rp2 from PC USB. |
| 2 | Prepare rp2 no. 2 | Connect rp2 to PC USB. Start VSCode. Edit config.py & main.py. Upload project to rp2. Disconnect rp2 from PC USB. |
| 3 | Start node 0 | Start master node at PC using the launch command: launch mp_async_example1_fedd_mean.py 3 0-0 0 192.168.2.4 |
| 4 | Start node 1 | Power on rp2 no. 1 (by a separate power unit) or connect rp2 no. 1 to PC and start main.py from VSCode (see the note below). |
| 5 | Start node 2 | Power on rp2 no. 2 (by a separate power unit) or connect rp2 no. 2 to PC and start main.py from VSCode (see the note below). |
| 6 | Verify results | Manually verify results by comparing them with the expected. See the note below. |

Note for verifying results for all the examples in this paper: For the node 0 the results are available in the launch terminal, whereas for the nodes 1 and 2 they are available in the VSCode terminal. However, only one rp2 can be controlled by a PC with the VSCode at a single experiment, so the experiment needs to be either conducted twice or the two PCs with the VSCode can be used in the single experiment (the latter approach should be used if the experiment conditions might change i.e., if we suspect that there are possible sources of nondeterminism e.g., possible hardware/software malfunctioning).

*4.2. Example 2: Centralized data averaging*

This example is the adapted Example 2 from [2]. Its purpose is the centralized averaging of the client models, and its pseudocode is given in Table 4 (when compared with [2] the new or modified lines are the lines 1, 3, 4, and 8).

Table 4
Example 2 pseudocode.

```
01: example2(noNodes, nodeId, flSrvId, mrIpAdr)
02:   // Create PtbFla object and await the system startup
03:   ptb = PtbFla(noNodes, nodeId, flSrvId, mrIpAdr)
04:   await ptb.start()
05:   // Set localData for FL server/clients as follows
06:   localData = [nodeId+1]
07:   // Call fl_centralized with noIterations = 10
08:   ret = await ptb.fl_centralized(servercb, clientcb, localData, None, 10)
09: clientcb(localData, privateData, msg)
10:   return [(localData[0] + msg[0])/2]
11: servercb(privateData, msgs)
12:   tmp = 0.0
13:   for lst in msgs:
14:     tmp = tmp + lst[0]
15:   tmp = tmp / len(msgs)
16:   return [tmp]
```

The main functions of Example2: (i) creates the object *ptb* and awaits the system startup, (ii) sets the initial local data of an instance to [*nodeId*+1], and (iii) awaits the coroutine fl_centralized on *ptb*. The initial local data in this example is a simple model that is encoded as a list with a single coefficient that characterizes client behaviour (e.g., an average value of some variable). In FL, the server model is more authoritative than the clients' models.

The client callback function clientcb averages this client model and the server model received through the argument *msg* i.e., it returns the list with the coefficient that is the average of the coefficients from this client list and the list in *msg*.

The server callback function servercb averages all the client models, which it receives through the argument *msgs* i.e., it returns the list with the coefficient that is the average of the coefficients of all the lists in *msgs*.

As shown in [2], the local data models, i.e., the coefficients in the lists are converging through the iterations to the average value 1.75, which is not the simple average of the initial coefficients of instances 1, 2, and 3 (that is 2), because the model at the first instance (the server) is more authoritative than the models at the other two (the clients). Therefore, the average value 1.75 is somewhat closer to the server's initial value 1, see [2] for more details.

The system configuration for the experimental validation is given in Table 5. Note that the last argument '192.168.2.4' was the specific IP address of the master node.

Table 5
Example 2 system configuration.

| Node index | Hosted by | Argument vector (argv) for the node |
|---|---|---|
| 0 | PC | ['mp_async_example2_cent_avg.py', '3', '0', '0', '192.168.2.4'] |
| 1 | Pico board | ['mp_async_example2_cent_avg.py', '3', '1', '0', '192.168.2.4'] |
| 2 | Pico board | ['mp_async_example2_cent_avg.py', '3', '2', '0', '192.168.2.4'] |

The operational instructions for the experimental validation are given in Table 6. The module main should import the Python module "mp_async_example2_cent_avg.py". The expected results may be provided beforehand by running all the instances on the PC using this launch command "launch mp_async_example2_cent_avg.py 3 id 0 192.168.2.4" – the results for all the three nodes will be displayed in three separate terminals, respectively.

Table 6
Example 2 operational instructions.

| Step no. | Step | Description (comment) |
|---|---|---|
| 1 | Prepare rp2 no. 1 | Connect rp2 no. 1 to PC over USB. Start VSCode. Edit config.py. Upload project to rp2 no. 1. Disconnect rp2 no. 1 from PC. |
| 2 | Prepare rp2 no. 2 | Connect rp2 no. 2 to PC over USB. Start VSCode. Edit config.py. Upload project to rp2 no. 2. Disconnect rp2 no. 2 from PC. |

| 3 | Start node 0 | Start master node at PC using the launch command: launch mp_async_example2_cent_avg.py 3 0-0 0 192.168.2.4 |
| 4 | Start node 1 | Power on rp2 no. 1 (by a separate power unit) or connect rp2 no. 1 to PC and start main.py from VSCode. |
| 5 | Start node 2 | Power on rp2 no. 2 (by a separate power unit) or connect rp2 no. 2 to PC and start main.py from VSCode. |
| 6 | Verify results | Manually verify results by comparing them with the expected. See the note at the end of Section 4.1. |

*4.3. Example 3: Decentralized data averaging*

This example is the adapted Example 3 from [2]. Its purpose is the decentralized averaging of the client models, and its pseudocode is given in Table 7 (when compared with [2] the new or modified lines are the lines 1, 3, 4, and 8).

Table 7
Example 3 pseudocode.

| |
| --- |
| 01: example3(*noNodes*, *nodeId*, *mrIpAdr*) |
| 02:   // Create PtbFla object and await the system startup |
| 03:   *ptb* = PtbFla(*noNodes*, *nodeId*, 0, *mrIpAdr*) |
| 04:   await *ptb*.start() |
| 05:   // Set localData for FL server/clients as follows |
| 06:   *localData* = [*nodeId*+1] |
| 07:   // Call fl_centralized with noIterations = 3 |
| 08:   *ret* = await *ptb*.fl_decentralized(servercb, clientcb, *localData*, None, 3) |
| 09: clientcb(*localData*, *privateData*, *msg*) |
| 10:   return [(*localData*[0] + *msg*[0])/2] |
| 11: servercb(*privateData*, *msgs*) |
| 12:   *tmp* = 0.0 |
| 13:   for *lst* in *msgs*: |
| 14:     *tmp* = *tmp* + *lst*[0] |
| 15:   *tmp* = *tmp* / len(*msgs*) |
| 16:   return [*tmp*] |

The pseudo code for this example is practically identical as for Example 2.

The main difference in the pseudo code for this example is in line 8, where the coroutine fl_decentralized is awaited instead of the coroutine fl_centralized. The other difference is that the variable *flSrvId* is not used by the fl_decentralized, therefore the lines 1 and 3 are different (*flSrvId* is set to 0 in the line 3, only because the constructor requires some value for this argument). Note that the callback functions (lines 9-16) are identical, but here they are called from the API coroutine fl_decentralized, so the overall behaviour is of course different.

As shown in [2], the local data models, i.e., the coefficients in the lists converge through the iterations to the average value 2.0, which is the simple average of the initial coefficients of instances 1, 2, and 3 (that is 2), because the models at all the instances are of equal authority (all the instances are equal peers). Interestingly, the decentralized algorithm converges faster than the centralized on – the former converges in the third iteration, whereas the latter converges in the tenth iteration, see [2] for more details.

The system configuration for the experimental validation is given in Table 8. Note that the last argument '192.168.2.4' was the specific IP address of the master node.

Table 8
Example 3 system configuration.

| Node index | Hosted by | Argument vector (argv) for the node |
| --- | --- | --- |
| 0 | PC | ['mp_async_example3_decent_avg.py', '3', '0', '192.168.2.4'] |
| 1 | Pico board | ['mp_async_example3_decent_avg.py', '3', '1', '192.168.2.4'] |
| 2 | Pico board | ['mp_async_example3_decent_avg.py', '3', '2', '192.168.2.4'] |

The operational instructions for the experimental validation are given in Table 9. The module main should import the python module "mp_async_example3_decent_avg.py". The expected results may be provided beforehand by running all the instances on the PC using this launch command "launch mp_async_example3_decent_avg.py.py 3 id 192.168.2.4" – the results for all the three nodes will be displayed in the three separate terminals, respectively.

Table 9
Example 3 operational instructions.

| Step no. | Step | Description (comment) |
|---|---|---|
| 1 | Prepare rp2 no. 1 | Connect rp2 no. 1 to PC over USB. Start VSCode. Edit config.py. Upload project to rp2 no. 1. Disconnect rp2 no. 1 from PC. |
| 2 | Prepare rp2 no. 2 | Connect rp2 no. 2 to PC over USB. Start VSCode. Edit config.py. Upload project to rp2 no. 2. Disconnect rp2 no. 2 from PC. |
| 3 | Start node 0 | Start master node at PC using the launch command: launch mp_async_example3_decent_avg.py.py 3 0-0 0 192.168.2.4 |
| 4 | Start node 1 | Power on rp2 no. 1 (by a separate power unit) or connect rp2 no. 1 to PC and start main.py from VSCode. |
| 5 | Start node 2 | Power on rp2 no. 2 (by a separate power unit) or connect rp2 no. 2 to PC and start main.py from VSCode. |
| 6 | Verify results | Manually verify results by comparing them with the expected. See the note at the end of Section 4.1. |

*4.4. Example 4: ODTS*

This new example is presented in this paper for the first time. The main purpose of this simplified ODTS simulation is to validate the coroutine get1Meas, and its pseudocode is given in Table 10. In this example, the nodes are equal peers (corresponding to individual SVs), and since they communicate in pairs (think of a constellation of SVs) we assume that the number of nodes is even (in the experiments it is set to 4).

The number of iterations to be executed is equal to the number of the time slot blocks (*noBlocks*, in the experiments set to 1) to be processed, whereas the size of the block is equal to the number of time slots (*noTSlots*, in the experiments set to 3) in the block. The ODTS simulation relies on the schedule of pairwise SV communications, which is here encoded as a dictionary dubbed *connections* with key-value pair, where the key is the pair (*tSlot*, *nodeId*) and the value is *peerId*, where *tSlot* is the time slot, *nodeId* is the ID of this node, and *peerId* is the ID of the peer node. This dictionary is constructed by the function islScheduling (not shown in the pseudocode because it is out of the scope of this paper).

The kernel of the ODTS distributed algorithm are the two nested loops, where the outer loop goes over block (i.e., iterations), the inner loop goes over time slots, and within the inner loop this node exchanges orbital data (*odata*) with its peer and calls the function odts to determine the new state. In this simplified ODTS simulation the orbital data of the node is just a single float, and the function odts just averages the orbital data of this node and the peer node, which is very different from odts function used in the real ODTS but is simple enough and good enough for the purpose of the get1Meas validation.

Table 10
Example 4 pseudocode.

```
01: example4(noNodes, nodeId, noBlocks, noTSlots, mrIpAdr)
02:   // Create PtbFla object and await the system startup
03:   ptb = PtbFla(noNodes, nodeId, 0, mrIpAdr)
04:   await ptb.start()
05:   // Set the odata for this node
06:   odata = 1.0 + nodeId
07:   // Set the initial state
08:   state = odata
09:   // Set the communication schedule
```

```
10:   connections = islScheduling(noNodes, noTSlots)
11:   // The ODTS kernel: Iterate over blocks and time slots
12:   for block in [0, noBlocks)
13:     for tSlot in [0, noTSlots)
14:       // Get peer ID
15:       peerId = connections[(tSlot, nodeId)]
16:       // Await get1Meas to get peer data
17:       obs = await ptb.get1Meas(peerId, odata)
18:       // Perform ODTS calculation and set the new state
19:       state = odts(state, obs)
20: // Simplified ODTS calculation
21: odts(state, obs)
22:   return (state + obs)/2.0
```

After the introductory notes made above, we explain the pseudocode in Table 10. The main function example4: (i) creates the object *ptb* and awaits the system startup, (ii) sets the orbital data of an instance to (1.0+*nodeId*), (iii) sets the initial state to *odata*, (iv) sets the communication schedule stored in the dictionary *connections*, and (v) performs the ODTS kernel comprising the two nested for loops.

The body of the inner loop: (i) gets the peer ID by reading the value associated with the tuple (*tSlot*, *nodeId*), (ii) gets the peer data by awaiting the coroutine get1Meas on *ptb*, and (iii) performs the ODTS calculation and sets the new state.

The function odts calculates the new state value by averaging the current state value and the data received from the peer. In the state-of-the-art ODTS typically the extended Calman filter is used to minimize the covariance of state vectors (SV's position and velocity) and in the TaRDIS project roadmap various ML based algorithms will be developed and compared.

The system configuration for the experimental validation is given in Table 11. Note that the last argument '192.168.2.4' was the specific IP address of the master node.

Table 11
Example 4 system configuration.

| Node index | Hosted by  | Argument vector (argv) for the node |
|------------|------------|-------------------------------------|
| 0          | PC         | ['mp_async_example6_odts.py', '4', '0', '1', '3', '192.168.2.4'] |
| 1          | Pico board | ['mp_async_example6_odts.py', '4', '1', '1', '3', '192.168.2.4'] |
| 2          | Pico board | ['mp_async_example6_odts.py', '4', '2', '1', '3', '192.168.2.4'] |
| 3          | PC         | ['mp_async_example6_odts.py', '4', '3', '1', '3', '192.168.2.4'] |

The operational instructions for the experimental validation are given in Table 12. The module main should import the python module "mp_async_example6_odts.py". The expected results may be provided beforehand by running all the instances on the PC using this launch command "launch mp_async_example6_odts.py 4 id 1 3 192.168.2.4" – the results for all the four nodes will be displayed in the four separate terminals, respectively. The final state for the nodes 0, 1, 2, and 3, should be the values 3.125, 2.375, 2.625, and 1.875, respectively.

Table 12
Example 4 operational instructions.

| Step no. | Step            | Description (comment) |
|----------|-----------------|------------------------|
| 1        | Prepare rp2 no. 1 | Connect rp2 no. 1 to PC over USB. Start VSCode. Edit config.py. Upload project to rp2 no. 1. Disconnect rp2 no. 1 from PC. |
| 2        | Prepare rp2 no. 2 | Connect rp2 no. 2 to PC over USB. Start VSCode. Edit config.py. Upload project to rp2 no. 2. Disconnect rp2 no. 2 from PC. |
| 3        | Start node 0    | Start master node at PC using the launch command: launch mp_async_example6_odts.py 4 0-0 1 3 192.168.2.4 |
| 4        | Start node 3    | Start master node at PC using the launch command: launch mp_async_example6_odts.py 4 3-3 1 3 192.168.2.4 |
| 5        | Start node 1    | Power on rp2 no. 1 (by a separate power unit) or connect rp2 no. |
| | | 1 to PC and start main.py from VSCode. |
| 6 | Start node 2 | Power on rp2 no. 2 (by a separate power unit) or connect rp2 no. 2 to PC and start main.py from VSCode. |
| 7 | Verify results | Manually verify results by comparing them with the expected. See the note at the end of Section 4.1. |

*4.5. Discussion*

As a short summary at the end of this section, we may conclude that MPT-FLA was successfully validated as the functional correctness of the four adapted algorithms was proved by the experiments, where the meaning of the functional correctness is producing the same numerical results. It seems that this should be sufficient to be taken as the proof of concept for the current MPT-FLA version.

However, it should be mentioned that we experienced some difficulties i.e., issues during the experiments that might compromise their validity. All of them seem to be related to the WiFi network.

The first issue is related to the Pico board attempts to connect to the WiFi network. At the beginning of a session, these attempts were always successful – in the case of our setup the number of these initial repetitive attempts ranged say up to 5. However, the subsequent attempts were becoming increasingly problematic in the sense that they were taking progressively more time. A possible reason for this issue is that the router treated these repetitive attempts as the denial of service (DoS) attacks. Therefore, we needed to make pauses between the individual sessions, and what is more problematic, we could not measure software reliability.

The second issue is related to the WiFi interferences, which happen both in the isolated WiFi networks and in the overlapping WiFi networks where they tend to become more severe. Under these conditions, TCP connections become more vulnerable, and the TCP exponential backoff causes progressively increasing network latencies. In extreme cases, when the experiments are conducted in apartment buildings, the communication could be completely broken. Therefore, we conducted the experiments in the laboratory where the interferences were less severe and in these almost ideal conditions, all the experiments were successful.

Related to the second issue, it is important to mention that in the case when we tried to make experiments in apartment buildings, it was not clear whether the communication crashes were caused by WiFi routers or the TCP sockets in the current MicroPython version, or something else.

**5. Conclusion**

This paper presents MicroPyton Testbed for Federated Learning Algorithms (MPT-FLA), a new FL framework that inherits all its predecessor PTB-FLA framework advantages and additionally allows individual application instances to run on different network nodes like PCs and IoTs, primarily in edge systems. The paper also presents the MPT-FLA experimental validation that was conducted on the appropriate experimental setup in the laboratory, by using the four application examples (the federated map, the centralized data averaging, the decentralized data averaging, and the ODTS).

The main original paper contributions are: (1) the novel FL framework dubbed MPT-FLA, which is based on MicroPython and Python asyncio, (2) a new set of application examples, which are adapted to the new MPT-FLA API, and (3) the experimental validation approach, results, and discussion, which may be useful to other researchers.

The main MPT-FLA advantages over PTB-FLA are: (1) it supports distributed applications whose instances may run on different network nodes like PCs and IoTs, whereas PTB-FLA requires all the instances to run in the single node (localhost), and (2) being based on the MicroPython and Python asyncio, it is a great match for IoTs based on smaller platforms like RPi Pico W boards, in contrast to PTB-FLA which is based on full Python and Python multiprocessing and as such is suited only for bigger platforms like PCs.

The main limitations of the research presented in this paper is that the experimental validation was made under ideal conditions in the laboratory by proving only the functional correctness of the simple application examples.

In the future work we plan to develop some benchmark applications and conduct more detailed MPT-FLA performance evaluation.

**Acknowledgements**

EU Funded by the European Union (TaRDIS, 101093006). Views and opinions expressed are however those of the author(s) only and do not necessarily reflect those of the European Union. Neither the European Union nor the granting authority can be held responsible for them.

**References**

[1] TaRDIS: Trustworthy and Resilient Decentralised Intelligence For Edge Systems. https://www.project-tardis.eu/, 2023 (accessed 1 February 2024)
[2] M. Popovic, M. Popovic, I. Kastelan, M. Djukic, and S. Ghilezan, A Simple Python Testbed for Federated Learning Algorithms, 2023, in: Proceedings of the 2023 Zooming Innovation in Consumer Technologies Conference, 2023, pp. 148-153, https://doi.org/10.1109/ZINC58345.2023.10173859.
[3] I. Prokić, S. Ghilezan, S. Kašterović, M. Popovic, M. Popovic, I. Kaštelan, Correct orchestration of Federated Learning generic algorithms: formalisation and verification in CSP, in: J. Kofron, T. Margaria, C. Seceleanu (Eds.), Engineering of Computer-Based Systems, Lecture Notes in Computer Science, Vol. 14390, Springer, Cham, 2024, pp. 274–288, https://doi.org/10.1007/978-3-031-49252-5_25.
[4] M. Popovic, M. Popovic, I. Kastelan, M. Djukic, and I. Basicevic, A Federated Learning Algorithms Development Paradigm, in: J. Kofron, T. Margaria, C. Seceleanu (Eds.), Engineering of Computer-Based Systems, Lecture Notes in Computer Science, Vol. 14390, Springer, Cham, 2024, pp. 26–41, https://doi.org/10.1007/978-3-031-49252-5_4.
[5] M. Popovic, M. Popovic, I. Kastelan, M. Djukic, I. Basicevic, Developing Elementary Federated Learning Algorithms Leveraging the ChatGPT, in: Proceedings of the 31st Telecommunications Forum (TELFOR 2023), IEEE Xplore, 2023, pp. 1-4, https://doi.org/10.1109/TELFOR59449.2023.10372714.
[6] H.B. McMahan, E. Moore, D. Ramage, S. Hampson, B.A. Arcas, Communication-Efficient Learning of Deep Networks from Decentralized Data, in Proceedings of the 20th International Conference on Artificial Intelligence and Statistics, 2017, pp. 1-10, URL https://proceedings.mlr.press/v54/mcmahan17a/mcmahan17a.pdf.
[7] TensorFlow Federated: Machine Learning on Decentralized Data. https://www.tensorflow.org/federated (accessed 1 February 2024).
[8] B. Ying, K. Yuan, H. Hu, Y. Chen, W. Yin, BlueFog: Make Decentralized Algorithms Practical for Optimization and Deep Learning, https://arxiv.org/abs/2111.04287, 2021 (accessed 1 February 2024).
[9] I. Kholod, E. Yanaki, D. Fomichev, E. Shalugin, E. Novikova, E. Filippov, M. Nordlund, Open-Source Federated Learning Frameworks for IoT: A Comparative Review and Analysis, Sensors 21 (1) (2021) 167, https://doi.org/10.3390/s21010167.
[10] A. Feraudo, P. Yadav, V. Safronov, D.A. Popescu, R. Mortier, S. Wang, P. Bellavista, J. Crowcroft, CoLearn: Enabling Federated Learning in MUD-compliant IoT Edge Networks, in: Proceedings of the 3rd International Workshop on Edge Systems, Analytics and Networking, 2020, pp. 25–30, https://doi.org/10.1145/3378679.3394528.
[11] T. Zhang, C. He, T. Ma, L. Gao, M. Ma, S. Avestimehr, Federated Learning for Internet of Things, in: Proceedings of the 19th ACM Conference on Embedded

Networked Sensor Systems, 2021, pp. 413–419, https://doi.org/10.1145/3485730.3493444.

[12] C. Shen, W. Xue, An Experiment Study on Federated Learning Testbed, in: Y.D. Zhang, T. Senjyu, C. So-In, A. Joshi, (Eds.), Smart Trends in Computing and Communications, Lecture Notes in Networks and Systems, Vol. 286, Springer, Singapore, 2022, pp. 209–217. https://doi.org/10.1007/978-981-16-4016-2_20.

[13] T.G. Mattson, B. Sanders, and B. Massingill, Patterns for Parallel Programming, Addison-Wesley, Massachusetts, USA, 2008.

[14] A. Shajii, G. Ramirez, H. Smajlović, J. Ray, B. Berger, S. Amarasinghe, I. Numanagić, Codon: A Compiler for High-Performance Pythonic Applications and DSLs, in Proceedings of the ACM SIGPLAN 2023 International Conference on Compiler Construction (CC'23), 2023, pp. 191–202. https://doi.org/10.1145/3578360.3580275.

[15] J. Lange, N. Ng, B. Toninho, N. Yoshida, A Static Verification Framework for Message Passing in Go using Behavioural Types, in: 2018 ACM/IEEE 40th International Conference on Software Engineering, 2018, pp. 1137-1148, https://doi.org/10.1145/3180155.3180157.